# Casimir Effect, Hawking Radiation and Trace Anomaly

M.R.Setare*

Department of Physics,Sharif University of Technology Tehran-Iran

October 22, 2018

## Abstract

The Casimir energy for massless scalar field of two parallel conductor, in two dimensional Schwarzchild black hole background, with Dirichlet boundary conditions is calculated by making use of general properties of renormalized stress tensor. We show that vacuum expectation value of stress tensor can be obtain by Casimir effect, trace anomaly and Hawking radiation. Four-dimensional of this problem, by this method, is under progress by this author.

*E-mail:Mreza@physics.sharif.ac.ir

# 1 Introduction

The Casimir effect is one of the most interesting manifestations of nontrivial properties of the vacuum state in quantum field theory [1, 2]. Since its first prediction by Casimir in 1948 [3], this effect has been investigated for various cases of boundary geometries and various types of field[4, 5]. The Casimir effect can be viewed as a polarization of vacuum by boundary conditions. Another type of vacuum polarization arises in the case of external gravitational field. In the static situation, the disturbance of the quantum state induces vacuum energy and stress, but no particles are created. The creation of particles from the vacuum takes place due to the interaction with dynamical external constraints. For example the motion of a single reflecting boundary (mirror) can create particles [6], the creation of particles by time-dependent external gravitational field is another example of dynamical external constraints.

It has been shown [7, 8] that particle creation by black hole in four dimension is as a consequence of the Casimir effect for spherical shell. It has been shown that the only existence of the horizon and of the barrier in the effective potential is sufficient to compel the black hole to emit black-body radiation with temperature that exactly coincides with the standard result for Hawking radiation. In [8], the results for the accelerated-mirror have been used to prove above statement. To see more about relation between moving mirrors and black holes refer to [9]

Another relation between Casimir effect and Schwarzschild black hole thermodynamic is thermodynamic instability. Widom et al [10, 11] showed the black hole capacity is negative, then an increase in its energy decreases its temperature. They showed that the electrodynamic Casimir effect can also produce thermodynamic instability.

In this paper the Casimir energy for massless scalar field in two dimensional Schwarzschild black hole for two parallel conducting plates with Dirichlet boundary conditions is calculated. The Casimir energy is obtained by imposing general requirements. Calculation of renormalized stress tensor for massless scalar field in two-dimensional Schwarzschild background has been done in [12] (for a pedagogical review see [13, 14]). Knowing of Casimir energy in flat space and trace anomaly help us to calculate renormalized stress tensor. To see similar calculation in background of static domain wall refer to [15]. In other situation, besides two previous quantities, we use Hawking radiation, which also has a contribution in stress tensor. Therefore the renormalized stress tensor is not unique and depends on the vacuum under consideration. Our paper is organised as follows: In section 2 general properties of stress tensor are discussed. Then in section 3, vacuum expectation value of stress tensor in two dimensions is obtained. Finally in section 4, we conclude and summarize the results.

# 2 General properties of stress tensor

In semiclassical framework for yielding a sensible theory of back reaction Wald [16] has developed an axiomatic approach. There one tries to obtain an expression for the renormalized $T_{\mu\nu}$ from the properties (axioms) which it must fulfill. The axioms for the renormalized energy momentum tensor are as follow:

1-For off-diagonal elements, standard result should be obtained.
2-In Minkowski space-time, standard result should be obtained.
3-Expectation values of energy momentum are conserved.
4-Causality holds .
5-Energy momentum tensor contains no local curvature tensor depending on derivatives of the metric higher than second order.

Two prescriptions that satisfy the first four axioms can differ at most by a conserved local curvature term. Wald, [17], showed any prescription for renormalized $T_{\mu\nu}$ which is consistent with axioms 1-4 must yield the given trace up to the addition of the trace of conserved local curvature. It must be noted that trace anomalies in stress-tensor, (that is, the nonvanishing $T^\mu_\mu$ for a conformally invariant field after renormalization) are originated from some quantum behavior [18]. In two dimensional space time one can show that a trace-free stress tensor can not be consistent with conservation and causality if particle creation occurs. A trace-free, conserved stress tensor in two dimensions must always remain zero if it is initially zero. One can show that the "Davies-Fulling-Unruh" formula [19] for the stress tensor of scalar field which yields an anomalous trace , $T^\mu_\mu = \frac{R}{24\pi}$, is the unique one which is consistent with the above axioms. In four dimensions, just as in two dimensions, a trace-free stress tensor which agree with the formal expression for the matrix elements between orthogonal states can not be compatible with both conservation laws and causality . It must be noted that, as Wald showed [17], with Hadamard regularization in massless case axiom (5) can not be satisfied unless we introduce a new fundamental length scale for nature. Regarding all these axioms, thus we are able to get an unambiguous prescription for calculation of stress tensor.

## 3 Vacuum expectation values of stress tensor

Our background shows a Schwarzschild black hole with following metric

$$d^2s = -(1 - 2\frac{m}{r})dt^2 + (1 - 2\frac{m}{r})^{-1}dr^2 + r^2(d\theta^2 + \sin^2 d\varphi^2). \tag{1}$$

Now we reduce dimension of spacetime to two,

$$d^2s = -(1 - \frac{2m}{r})dt^2 + (1 - \frac{2m}{r})^{-1}dr^2. \tag{2}$$

The metric (2) can be written in conformal form

$$d^2s = \Omega(r)(-dt^2 + dr^{*2}), \tag{3}$$

with

$$\Omega(r) = 1 - \frac{2m}{r}, \qquad \frac{dr}{dr^*} = \Omega(r). \tag{4}$$

From now on, our main goal is the determination of a general form of conserved energy momentum tensor with regarding trace anomaly for the metric (2). For the non zero Christoffel symbols of the metric (2) we have in $(t, r^*)$ coordinates:

$$\Gamma^{r^*}_{tt} = \Gamma^t_{tr^*} = \Gamma^t_{r^*r^*} = \Gamma^{r^*}_{r^*r^*} = \frac{1}{2}\frac{d\Omega(r)}{dr}. \tag{5}$$

Then the conservation equation takes the form

$$\partial_{r^*}T^{r^*}_t + \Gamma^t_{tr^*}T^{r^*}_t - \Gamma^{r^*}_{tt}T^t_{r^*} = 0, \tag{6}$$

$$\partial_{r^*}T^{r^*}_{r^*} + \Gamma^t_{tr^*}T^{r^*}_{r^*} - \Gamma^t_{tr^*}T^t_t = 0, \tag{7}$$

in which

$$T^t_{r^*} = -T^{r^*}_t, \qquad T^t_t = T^\alpha_\alpha - T^{r^*}_{r^*} \tag{8}$$

and $T^\alpha_\alpha$ is anomalous trace in two dimension. Using equations $(5 - 8)$ it can be shown that

$$\frac{d}{dr}(\Omega(r)T^{r^*}_t) = 0 \tag{9}$$

and

$$\frac{d}{dr}(\Omega(r)T^{r^*}_{r^*}) = \frac{1}{2}(\frac{d\Omega(r)}{dr})T^\alpha_\alpha. \tag{10}$$

Then equation (9) leads to
$$T_t^{r^*} = \alpha \Omega^{-1}(r), \tag{11}$$
where $\alpha$ is a constant of integration. The solution of Eq.(10) may be written in the following form
$$T_{r^*}^{r^*}(r) = (H(r) + \beta)\Omega^{-1}(r), \tag{12}$$
where
$$H(r) = 1/2 \int_l^r T_\alpha^\alpha(r') \frac{d}{dr'} \Omega(r') dr', \tag{13}$$
with $l$ being an arbitrary scale of length and considering
$$T_\alpha^\alpha = \frac{R}{24\pi} = \frac{m}{6\pi r^3}. \tag{14}$$

The function $H(r)$ produces the non-local contribution of the trace $T_\alpha^\alpha(x)$ to the energy momentum tensor. Given a length scale $l$, one can show that[20],[6]
$$l \approx r_b = 2m, \tag{15}$$
$$H(r) = \frac{m^2}{24\pi} \left( \frac{1}{16m^4} - \frac{1}{r^4} \right). \tag{16}$$

Using the equations (8), (11) and (12) it can be shown that in $(t, r^*)$ coordinates, we have the following most general form for the stress tensor in our interested background
$$T^\mu_{\ \nu}(r) = \begin{pmatrix} T_\alpha^\alpha - \Omega(r)^{-1} H(r) & 0 \\ 0 & \Omega(r)^{-1} H(r) \end{pmatrix} + \Omega^{-1} \begin{pmatrix} -\beta & -\alpha \\ \alpha & \beta \end{pmatrix}. \tag{17}$$

We will obtain two constants $\alpha$ and $\beta$ by imposing the second axiom of renormalization scheme. If we put $m = 0$, then we reach the special case of Minkowski space-time. Here we introduce the state vector $|C>$ which is analog for the Boulware state [21]. In the case of existence of boundary the Minkowski limt of $|C>$ is not the Minkowski state $|M>$. In this limit $<C|T_{\mu\nu}|C>$ is non zero and shows the effects of boundary conditions on vacuum of scalar filed. We have chosen the Dirichlet boundary condition $\phi(r_1) = \phi(r_2) = 0$. The standard regularized casimir stress tensor in Minkowskian space-time is as follows
$$reg < 0|T_{\mu\nu}|0> = \frac{-\pi}{24a^2} \delta_{\mu\nu}, \tag{18}$$

in which $a$ is the proper distance between the plates. We can obtain it using mode summation with the help, for example, Abel-Plana summation formula [22, 2] without introducing the cut off. Comparing to (17) we obtain
$$\beta = \frac{\pi}{24a^2} - \frac{1}{384\pi m^2}, \qquad \alpha = 0. \tag{19}$$

Now we consider the Hartle-Hawking state $|H>$ [23]. This state is not empty at infinity, even in absence of boundary conditions on quantum filed, corresponds to a thermal distribution of quanta at the Hawking temperature $T = \frac{1}{8\pi m}$. In fact, the state $|H>$ is related to a black hole in equilibrium with an infinite reservoir of black body radiation. In absence of boundary conditions stress tensor at infinity is equal to
$$<H|T^\mu_{\ \nu}|H> = \frac{\pi T^2}{12} \begin{pmatrix} 2 & 0 \\ 0 & -2 \end{pmatrix}. \tag{20}$$

In the presence of the boundary conditions, (18) have to be add to the above relation. In this case at $r \to \infty$ we have

$$< H|T^\mu_\nu|H > = (\frac{1}{384\pi m^2} - \frac{\pi}{24a^2})\begin{pmatrix} 1 & 0 \\ 0 & -1 \end{pmatrix}, \tag{21}$$

therefore

$$\beta = \frac{\pi}{24a^2} - \frac{1}{192\pi m^2}, \qquad \alpha = 0. \tag{22}$$

Difference between (22) and (19) is the existence of the bath of thermal radiation at temperature $T$.

In order to calculate the contribution from the Hawking evaporation process to the Casimir energy(total vacuum energy), for this special geometry, we introduce the last quantum state which is a convenient candidate for vacuum [24]. This state is called as Unruh state $|U>$. In the limit $r \to \infty$, this state corresponds to the outgoing flux of a black body radiation in the black hole temperature $T$. The stress tensor in the limit $r \to \infty$ in the absence of the boundary conditions is in the form:

$$< U|T^\mu_\nu|U > = \frac{\pi T^2}{12}\begin{pmatrix} 1 & 1 \\ -1 & -1 \end{pmatrix}. \tag{23}$$

In the presence of the boundary conditions the expression (18) should be added to the relation(23) which is a new stress tensor. Comparing the new stress with (17) one obtains:

$$\alpha = \frac{1}{768\pi m^2}, \qquad \beta = \frac{\pi}{24a^2} - \frac{1}{256\pi m^2}. \tag{24}$$

Finally stress tensor(17) in Boulware ,Hartle-Hawking and Unruh states is

$$< B|T^\mu_\nu(r)|B > = \begin{pmatrix} T^\alpha_\alpha - \Omega^{-1}(r)H(r) & 0 \\ 0 & \Omega^{-1}H(r) \end{pmatrix} + \Omega^{-1}(\frac{-\pi}{24a^2} + \frac{1}{384\pi m^2})\begin{pmatrix} 1 & 0 \\ 0 & -1 \end{pmatrix}. \tag{25}$$

$$< H|T^\mu_\nu(r)|H > = \begin{pmatrix} T^\alpha_\alpha - \Omega^{-1}(r)H(r) & 0 \\ 0 & \Omega^{-1}H(r) \end{pmatrix} + \Omega^{-1}(\frac{-\pi}{24a^2} + \frac{1}{192\pi m^2})\begin{pmatrix} 1 & 0 \\ 0 & -1 \end{pmatrix}. \tag{26}$$

$$< U|T^\mu_\nu(r)|U > = \begin{pmatrix} T^\alpha_\alpha - \Omega^{-1}(r)H(r) & 0 \\ 0 & \Omega^{-1}H(r) \end{pmatrix} + \Omega^{-1}\begin{pmatrix} \frac{-\pi}{24a^2} + \frac{1}{256\pi m^2} & \frac{-1}{768\pi m^2} \\ \frac{1}{768\pi m^2} & \frac{\pi}{24a^2} - \frac{1}{256\pi m^2} \end{pmatrix} \tag{27}$$

. Here the presence of

$$\frac{\pi}{24a^2}\Omega^{-1}\begin{pmatrix} -1 & 0 \\ 0 & 1 \end{pmatrix}. \tag{28}$$

is due to the boundary conditions. Therefore (25),(26) and (27) are separable as follows:

$$< B|T^\mu_\nu|B > = < B|T^{(g)\mu}_\nu|B > + < B|T^{(b)\mu}_\nu|B >. \tag{29}$$

$$< H|T^\mu_\nu|H > = < H|T^{(g)\mu}_\nu|H > + < H|T^{(b)\mu}_\nu|H > + < H|T^{(t)\mu}_\nu|H >. \tag{30}$$

$$< U|T^\mu_\nu|U > = < U|T^{(g)\mu}_\nu|U > + < U|T^{(b)\mu}_\nu|U > + < U|T^{(r)\mu}_\nu|U >. \tag{31}$$

where $< T^{(g)\mu}_\nu >$ stands for gravitational, $< T^{(b)\mu}_\nu >$ stands for boundary part, $< T^{(t)\mu}_\nu >$ corresponds to the bath of thermal radiation at temperature $T$, and $< T^{(r)\mu}_\nu >$ is Hawking radiation contribution. It should be noted that trace anomaly has contribution in the first term only which comes from the background not boundary effect. However it has contribution to the total Casimir energy momentum tensor. In the regions $r < r_1$ and $r > r_2$ the boundary part is zero and only gravitational polarization part is present.
The vacuum boundary part pressures acting on plates are

$$P_b^{(1,2)} = P_b(r = r_{1,2}) = - <T_1^{(b)1}(r = r_{1,2})> = -\Omega^{-1}(r_{1,2})\frac{\pi}{24a^2}, \qquad (32)$$

and have attractive nature. The effective pressure created by other parts in (25-27) are the same from the both sides on the plates, and hence leads to the zero effective force.

## 4 Conclusion

We have found the renormalized energy-momentum tensor for massless scalar field on background of 1+1 dimensional Schwarzschild black hole for two parallel plates with Dirichlet boundary conditions, by making use of general properties of stress tensor only. We propose that if we know the stress tensor for a given boundary in Minkowski space-time, the Casimir effect in gravitational background can be calculated. We have found direct relation between trace anomaly and total Casimir energy. In addition, by considering the Hawking radiation for observer far from black hole, who is the same as Minkowski observer, this radiation contributes to the Casimir effect.

In this paper we have derived three renormalized energy-momentum tensors for our case of study. This is due to selecting three vacuum states for calculation. If we consider Boulware vacuum, stress tensor will have two parts: boundary part and gravitational part. But using Hartle-Hawking and Unruh vacuums will result in another term added to stress tensor, which respectively corresponds to a bath of thermal radiation and Hawking radiation.

It seems that similar results can be obtain for four-dimensional Schwarzschild black hole [25].

**Acknowledgement**
I would like to thank Prof A.Saharian for useful discussion, Prof D.V.Vassilevich for the reference [13] introduce me and Mr.A.Rezakhani for reading manuscript. .